\documentstyle[12pt]{article}

\begin{document}

\begin{center}

\baselineskip 40pt

\vskip 2cm

{\Large {\bf Dynamics of a two-level system coupled to a quantum
oscillator: Transformed rotating-wave approximation }}

\vskip 1cm

{\large {\rm Congjun Gan and Hang Zheng}}

Department of Physics, Shanghai Jiao Tong University \\
Shanghai 200240, People's Republic of China \\

{\bf Abstract}

\end{center}

\baselineskip 20pt

For studying the dynamics of a two-level system coupled to a quantum
oscillator we have presented an analytical approach, the transformed
rotating-wave approximation, which takes into account the effect of
the counter-rotating terms but still keeps the simple mathematical
structure of the ordinary rotating-wave approximation. We have
calculated the energy levels of ground and lower-lying excited
states, as well as the time-dependent quantum dynamics. It is
obvious that the approach is quite simple and can be easily extended
to more complicated situation. Besides, the results are compared
with the numerically exact ones to show that for weak and
intermediate coupling and moderate detuning our analytic
calculations are quantitatively in good agreement with the exact
ones.

\vskip 1cm


\pagebreak

\baselineskip 20pt

\section{Introduction}

The physics of a two-level system coupled to a quantum oscillator
(spin-oscillator model, SOM) is of wide interest because it provides
a simple but ubiquitous model for numerous physical processes, such
as the superconducting qubit of Josephson
junction\cite{sc1,sc2,sc3,sc4,scrwa,iri1}, the semiconductor quantum
dot\cite{semi,ima}, the coupling between a qubit and a
nanomechanical oscillator\cite{iri2,iri3}, and a toy model for
Holstein polaron\cite{sas}. The Hamiltonian of SOM reads
\begin{equation}
H={\frac{1}{2}}\Omega \sigma _{x}+\omega b^{\dag }b
+{\frac{g}{2}}(b^{\dag }+b)\sigma_z.
\end{equation}
$\Omega$ is the level difference and $\sigma _{x}$ and $\sigma _{z}$
are Pauli matrices to describe the two-level system. $b^{\dag }$
($b$) is the creation (annihilation) operator of the quantum
oscillator with frequency $\omega$ and $g$ the coupling constant
between the two-level system and the oscillator. The model seems
quite simple, however, an analytical solution has not yet been found
and various approximate analytical and numerical methods have been
used.

The Hamiltonian (1) is equivalent to the famous Jaynes-Cummings
model with inclusion of both the rotating-wave terms and the
counter-rotating terms\cite{jc,jcr},
\begin{equation}
H_{JC}=\frac{1}{2}\Omega\sigma _{z}+\omega b^{\dag
}b-\frac{g}{2}\sigma _{x}(b^{\dag }+b),
\end{equation}%
if a rotation around the y axis is taken for the Pauli matrices,
$e^{i\pi\sigma_y/4}\sigma_x e^{-i\pi\sigma_y/4}=\sigma_z$ and
$e^{i\pi\sigma_y/4}\sigma_z e^{-i\pi\sigma_y/4}=-\sigma_x$. Then,
$b^{\dag}$ ($b$) is the creation (annihilation) operator of the
cavity mode and $\Omega$ is the atomic transition frequency ($g$ is
the vacuum Rabi frequency). Becides, the Hamiltonian (1) is also of
interest in the research area of matter-field interaction because it
is the following atom-field interaction $H_{AF}$ in the rotating
frame\cite{ima,wzz},
\begin{equation}
H_{AF}={1\over 2}\omega_0\sigma_z+{\frac{1}{2}}\Omega \left(\sigma
_{+}e^{-i\omega_ft}+\sigma_{-}e^{i\omega_ft}\right)+\omega b^{\dag
}b +{\frac{g}{2}}(b^{\dag }+b)\sigma_z,
\end{equation}
where $\sigma_{\pm}={1\over 2}(\sigma_x\pm i\sigma_y)$. Here
$\omega_0$ is the atom transition frequency (or the exciton energy
in quantum dot) and $\omega_f=\omega_0$ is the frequency of laser
field (in resonance with $\omega_0$), and $\Omega$ the Rabi
frequency. The dynamical evolution of the interacting Hamiltonian
$H_{AF}$ can be described by
\begin{eqnarray}
&&P(t)=\langle \psi_{AM}(t)| \sigma
_{z}|\psi_{AM}(t)\rangle=\langle \psi(t)| \sigma
_{z}|\psi(t)\rangle,
\end{eqnarray}
where $|\psi_{AM}(t)\rangle$ is the wave function of $H_{AM}$ and
$|\psi(t)\rangle=e^{i\omega_ft\sigma_z/2}|\psi_{AM}(t)\rangle$ is
that of $H$ (Eq.(1)). The initial state of the system is assumed
to be
$|\psi(0)\rangle=|\psi_{AM}(0)\rangle=|\uparrow\rangle|0\rangle$,
where $\sigma_z|\uparrow\rangle=|\uparrow\rangle$ is the
eigenstate of $\sigma_z$ and $|0\rangle$ is the vacuum state of
the quantum oscillator. When the coupling $g=0$ it is easily to
get the typical Rabi oscillation $P(t)=\cos(\Delta t)$. For $g\ne
0$, the Rabi oscillation may be modulated by the interaction with
the quantum oscillator and it is an interesting problem related to
the quantum manipulation of the interacting
system\cite{iri1,iri3}.

Although there is still no analytically exact solution for (1) or
(2), various approximate analytical solutions for (1) or (2)
already exist\cite{iri2,iri3,mie,lev,ita,san}. The most popular
one may be the rotating-wave approximation (RWA), which relies on
the assumption of near resonance ($|\Omega-\omega|\ll
\Omega\mbox{~and~}\omega$) and weak coupling ($g\ll
\Omega\mbox{~and~}\omega$). However, the condition of near
resonance may not be satisfied if we start from Hamiltonian (3),
because $\Omega$ is the Rabi frequency of the pumping field which
may be much larger or smaller than the oscillator
frequency\cite{ima,wzz}. Besides, quantum-limited solid-state
devices are an alternative to the usual atom-cavity implementation
of the SOM, and in these solid-state devices the coupling strength
$g$ and the detuing $|\Omega-\omega|$ may be outside the regime
where the RWA is valid (the coupling between a nanomechenical
resonator and a charge qubit may be in the intermediate range
$g/\omega\approx 0.1\sim 1$\cite{iri1,iri3,san}). This is the
motivation for approximate analytical solutions beyond the RWA to
be presented in last years.

The Hamiltonian (1) or (2) can be numerically solved easily and
quickly by ordinary PC, then, why do we still need an approximate
analytical solution? As far as we can see, the purpose may be: 1.to
see more clearly the physics, 2.to test the accuracy of the
analytical solution for extending it to more complicated model where
a numerically exact solution is difficult to be obtained. Hence, we
have following criterion for the validity of an approximate
analytical solution: First, it should be as simple as possible,
then, it can be easily extended to more complicated situation.
Second, the main physics should be catched, at least for the
interesting and concerned range of the parameters, and it should be
as accurate as possible compared with the numerically exact result.

Recently, one of us proposed an analytic approach\cite{zhe1} to the
spin-boson model, which describes a two-level system coupled to a
dissipative environment\cite{rmp}. Roughly speaking, Hamiltonian (1)
is a simplified version of the spin-boson model, that is, the single
mode spin-boson model. Our approach, which is a perturbation
expansion based on the unitary transformation, has been successfully
applied to several problems related to the interaction between the
quantum-limited system and its environment\cite{group,zhe2,wzz}. In
this work, we would show the validity of our analytic approach by
studying the numerically solvable model, which leads to the
transformed rotating-wave approximation (transformed RWA) for model
(1). We will focus on the dynamical evolution $P(t)$, since the
accuracy of its calculation depends not only on the calculation of
ground state but also on that of the lower-lying excited states. The
dynamical evolution $P(t)$ calculated by our transformed RWA will be
compared to the numerically exact one, as well as to the ordinary
RWA, to show that for weak and intermediate coupling and moderate
detuning our calculations are quantitatively in good agreement with
the numerically exact results.

\section{Theoretical analysis}

The ordinary RWA for the SOM (Eq.(1)) is
\begin{equation}
H_{RWA}={\frac{1}{2}}\Omega \sigma _{x}+\omega b^{\dag }b
+{\frac{g}{2}}\left [ b ^{\dag}|s_1\rangle\langle
s_2|+b|s_2\rangle\langle s_1| \right],
\end{equation}
where $\sigma_x|s_1\rangle=-|s_1\rangle$ and
$\sigma_x|s_2\rangle=|s_2\rangle$ are the eigenstates of
$\sigma_x$. It is easily to see that $|s_1\rangle |0\rangle$
($|0\rangle$: the vacuum state of the oscillator) is the exact
ground state of $H_{RWA}$ and the mathematical structure of
$H_{RWA}$ is quite simple.

We apply a unitary transformation\cite{zhe1} to $H$ (Eq.(1)),
$H^{\prime }=\exp (S)H\exp (-S)$, and the purpose of the
transformation is to take into account the effect of
counter-rotating terms, where
\begin{equation}
S=\frac{g}{2\omega}\xi\sigma _{z}(b^{\dag }-b).
\end{equation}
A parameter $\xi$ is introduced in $S$ and its form will be
determined later. The transformation can be done to the end and
the result is
\begin{eqnarray}
&&H^{\prime }=H_{0}^{\prime }+H_{1}^{\prime }+H_{2}^{\prime }, \\
&&H_{0}^{\prime }={\frac{1}{2}}\eta \Omega \sigma _{x}+\omega b
^{\dag }b -\frac{g ^{2}}{4\omega }\xi (2-\xi ), \\
&&H_{1}^{\prime }={\frac{1}{2}}g (1-\xi  )(b ^{\dag }+b )\sigma
_{z} +{\frac{1}{2}}\eta \Omega i\sigma_y \frac{g }{\omega }
\xi  (b ^{\dag }-b ),  \\
&&H_{2}^{\prime }= {\frac{1}{2}}\Omega\sigma_x\left( \cosh \{
\frac{g }{\omega  }\xi  (b ^{\dag }-b )\}-\eta \right)\nonumber\\
&&+{\frac{1}{2}}\Omega i\sigma_y\left( \sinh \{  \frac{g }{ \omega
}\xi  (b ^{\dag }-b )\}-\eta   \frac{g }{ \omega
 }\xi  (b ^{\dag }-b )\right),
\end{eqnarray}
where
\begin{eqnarray}
&&\eta =\exp [-  \frac{g ^{2}}{2\omega  ^{2}}\xi  ^{2}].
\end{eqnarray}
Obviously, $H_{0}^{\prime}$ can be solved exactly because for
which the spin and the oscillator are decoupled. If the
displacement parameter $\xi$ is determined as
\begin{eqnarray}
&&\xi =\frac{\omega  }{\omega  +\eta\Omega},
\end{eqnarray}
then we have
\begin{eqnarray}
&&H'_{1}=\frac{1}{2}\eta\Omega \frac{g }{\omega  }\xi \left [ b
^{\dag}(\sigma_z+i\sigma_y)+b
(\sigma_z-i\sigma_y)\right]=\eta\Omega \frac{g }{\omega }\xi\left
[ b ^{\dag}|s_1\rangle\langle s_2|+b|s_2\rangle\langle s_1|
\right].
\end{eqnarray}
Note that $H'_{1}$ is of the same form as the RWA in Eq.(5), except
the different coefficient.

The transformed Hamiltonian $H'$ is equivalent to the original $H$
and there is no approximation till this point. In the following, the
transformed Hamiltonian is approximated as $H'\approx
H_{TRWA}=H'_{0}+H'_{1}$, since the terms in $H'_{2}$ are related to
the double- and multiple-boson transition and their contributions to
the physical quantities are $O(g^4 )$. That means, through the
unitary transformation approach we get the transformed RWA
Hamiltonian $H_{TRWA}$, which is of the same mathematical structure
as the ordinary RWA $H_{RWA}$ in (5). $|s_1\rangle |0\rangle$ is the
exact ground state of $H_{TRWA}$ with ground state energy
\begin{eqnarray}
&&E_g=-\frac{1}{2}\eta\Omega - \frac{g ^{2}}{4\omega  }\xi (2-\xi
 ).
\end{eqnarray}
The eigenenergy for excited states can be easily obtained from
$H'_1$ (Eq.(13)), since it contains the rotating-wave terms only.
For $n=0,1,2,...$, the eigenenegies for all excited states are
\begin{eqnarray}
&&E_{2n+1}=(n+{1\over 2})\omega-{1\over
2}\sqrt{(\omega-\eta\Omega)^2+g'^2(n+1)} - \frac{g ^{2}}{4\omega
}\xi (2-\xi), \\
&&E_{2n+2}=(n+{1\over 2})\omega+{1\over
2}\sqrt{(\omega-\eta\Omega)^2+g'^2(n+1)} - \frac{g ^{2}}{4\omega
}\xi (2-\xi),
\end{eqnarray}
where $g'=2g\eta\Omega/(\omega+\eta\Omega)$.

The calculations of the ground state and lower-lying excited states
with moderate detuning are shown in Figs.1 and 2. Fig.1 is for the
case of $\Omega=1$ and $\omega=0.5$ but Fig.2 for $\Omega=0.5$ and
$\omega=1$. For comparison, the results of exact diagonalization and
those of generalized RWA\cite{iri3} are also shown. For ground state
energy our result of transformed RWA is very close to the exact
result and is better than that of generalized RWA. For excited
states we could not say definitely which one is better, but we can
say that for intermediate coupling $g/\omega\approx 0.1\sim 1$ our
results of transformed RWA are quite close to the exact results.

\section{Dynamical evolution $P(t)$}

We have shown that for the weak and intermediate coupling our
calculation of the ground state and the lower-lying excited state is
quantitatively in good agreement with the numerically exact results,
which is a check of the validity of our approach. Furthermore, in
this section our approach will be checked by calculation of the
excited state properties, that is, the dynamical evolution $P(t)$ of
Eq.(4).

The numerical exact calculation for $P(t)$ with Hamiltonian $H$ can
be obtained by the numerical diagonalization. The following is the
calculation of $P(t)$ in our transformed RWA. From Eq.(4) we have
\begin{eqnarray}
&&P(t)=\langle \psi(0)|e^{iHt} \sigma _{z}e^{-iHt}|\psi(0)\rangle
=\langle \psi(0)|e^{-S}e^{iH't}e^{S} \sigma
_{z}e^{-S}e^{-iH't}e^{S}|\psi(0)\rangle\nonumber\\
&& \approx\langle \psi'(0)|\exp(iH_{TRWA}t) \sigma
_{z}\exp(-iH_{TRWA}t)|\psi'(0)\rangle\nonumber\\
&&=\langle \psi'_I(t)|e^{iH'_0t} \sigma
_{z}e^{-iH'_0t}|\psi'_I(t)\rangle,
\end{eqnarray}
where $|\psi'(0)\rangle=e^{S}|\psi(0)\rangle$. Here the unitary
transformations, $e^S He^{-S}=H'$ and
$e^S\sigma_ze^{-S}=\sigma_z$, have been used.
$|\psi'_I(t)\rangle=e^{iH'_0t}\exp(-iH_{TRWA}t)|\psi'(0)\rangle$
is the wave function in interaction picture, which is the solution
of following Schroedinger equation,
\begin{eqnarray}
&&i \frac{\partial }{\partial
t} |\psi'_I(t)\rangle=H'_1(t)|\psi'_I(t)\rangle ,\\
&&H'_1(t)=e^{iH'_{0}t}H'_{1}e^{-iH'_{0}t}=\frac{g'}{2} \left( b
^{\dag}|s_1\rangle\langle s_2| e^{i(\omega-\eta\Omega)
t}+b|s_2\rangle\langle s_1| e^{-i(\omega-\eta\Omega) t}\right).
\end{eqnarray}

The initial state is
$|\psi'(0)\rangle=e^S|\uparrow\rangle|0\rangle={1\over
\sqrt{2}}(|s_1\rangle+|s_2\rangle)e^{\alpha(b^{\dag}-b)}|0\rangle$
with $\alpha=g\xi/2\omega$. Here the exponential operator is
expanded to the order $\alpha$: $e^{\alpha(b^{\dag}-b)}\approx
1+\alpha(b^{\dag}-b)$ and, hence, the initial condition is:
\begin{eqnarray}
&&
C_{10}(0)=C_{20}(0)={1\over\sqrt{2}},\mbox{~~~}C_{11}(0)=C_{21}(0)
=\frac{\alpha}{\sqrt{2}},\mbox{~~~}C_{12}(0)=0,
\end{eqnarray}
with the wave function\cite{scu}
\begin{eqnarray}
&& |\psi'_I(t)\rangle=C_{10}(t)|s_1\rangle
|0\rangle+C_{20}(t)|s_2\rangle
|0\rangle\nonumber\\
&&+C_{11}(t)|s_1\rangle |1\rangle+C_{21}(t)|s_2\rangle
|1\rangle+C_{12}(t)|s_1\rangle |2\rangle.
\end{eqnarray}
The Schroedinger equation is\cite{scu}
\begin{eqnarray}
&&i \frac{d }{d t}C_{10}(t)=0,\\
&&i \frac{d }{d
t}C_{20}(t)=\frac{g'}{2}e^{i(\eta\Omega-\omega)t}C_{11}(t),\mbox{~~~}i
\frac{d }{d
t}C_{11}(t)=\frac{g'}{2}e^{-i(\eta\Omega-\omega)t}C_{20}(t),\\
&&i \frac{d }{d
t}C_{21}(t)=\frac{g'}{2}e^{i(\eta\Omega-\omega)t}C_{12}(t),\mbox{~~~}i
\frac{d }{d
t}C_{12}(t)=\frac{g'}{2}e^{-i(\eta\Omega-\omega)t}C_{21}(t).
\end{eqnarray}
Then, the solution for Eq.(23) is
\begin{eqnarray}
&&C_{20}(t)={1\over\sqrt{2}}\left\{\left[\cos\frac{\Phi_0t}{2}-i\frac{\eta
\Omega-\omega}{\Phi_0}\sin\frac{\Phi_0t}{2}\right]
-\alpha\frac{ig'}{\Phi_0}\sin\frac{\Phi_0t}{2}\right\}
e^{i(\eta\Omega-\omega)t/2},  \\
&&C_{11}(t)={1\over\sqrt{2}}\left\{\alpha\left[\cos\frac{\Phi_0t}{2}+i\frac{\eta
\Omega-\omega}{\Phi_0}\sin\frac{\Phi_0t}{2}\right]
-\frac{ig'}{\Phi_0}\sin\frac{\Phi_0t}{2}\right\}
e^{-i(\eta\Omega-\omega)t/2},
\end{eqnarray}
and the solution for Eq.(24) is
\begin{eqnarray}
&&C_{21}(t)=\frac{\alpha}{\sqrt{2}}\left[\cos\frac{\Phi_1t}{2}-i\frac{\eta
\Omega-\omega}{\Phi_1}\sin\frac{\Phi_1t}{2}\right]
e^{i(\eta\Omega-\omega)t/2},  \\
&&C_{12}(t)=-\frac{\alpha}{\sqrt{2}}\frac{i\sqrt{2}g'}{\Phi_1}\sin\frac{\Phi_1t}{2}
e^{-i(\eta\Omega-\omega)t/2}.
\end{eqnarray}
where $\Phi_0^{2}=(\eta\Omega-\omega)^{2}+g'^{2}$ and
$\Phi_1^{2}=(\eta\Omega-\omega)^{2}+2g'^{2}$.

The dynamical evolution is
\begin{eqnarray}
&&P(t)=2\mbox{Re}\left(C^*_{20}(t)C_{10}(t)e^{i\eta\Omega t}+
C^*_{21}(t)C_{11}(t)e^{i\eta\Omega t}\right).
\end{eqnarray}

For the ordinary RWA (Eq.(5)), the interaction is
\begin{eqnarray}
&&H_1(t)=\frac{g}{2} \left( b ^{\dag}|s_1\rangle\langle s_2|
e^{i(\omega-\Omega) t}+b|s_2\rangle\langle s_1| e^{-i(\omega-\Omega)
t}\right).
\end{eqnarray}
Then, it is easily to get the dynamical evolution,
\begin{eqnarray}
&&P_{RWA}(t)=\cos\frac{\Phi_{0,RWA}t}{2}\cos\frac{(\Omega+\omega)t}{2}-\frac{\Omega-\omega}
{\Phi_-}\sin\frac{\Phi_{0,RWA}t}{2}\sin\frac{(\Omega+\omega)t}{2},
\end{eqnarray}
where $\Phi_{0,RWA}^{2}=(\Omega-\omega)^{2}+g^{2}$.

The calculations of $P(t)$ with moderate detuning and intermediate
coupling are shown in Figs.3 and 4. Fig.3 is for the larger
$\Omega$ case ($\Omega=1$, $\omega=0.5$ and $g=0.4$) but Fig.4 for
larger $\omega$ case ($\Omega=0.5$, $\omega=1$, and $g=0.5$). For
comparison, the results of exact diagonalization and those of
ordinary RWA are also shown. one can see that our results of
transformed RWA are quite close to the numerically exact ones,
much better than those of ordinary RWA (especially for the case of
Fig.4).

For some special case the ordinary RWA is absolutely invalid.
Fig.5 shows the $P(t)$ dynamics for large detuning $\omega=1$,
$\Omega=0.25$ and strong conpling $g=1$. One can see that the
result of ordinary RWA is qualitatively incorrect, but that of our
transformed RWA is still in qualitative agreement with the
numerically exact one.

\section{Concluding remarks}

We have presented an analytical approach, the transformed RWA, for
the SOM, which takes into account the effect of counter-rotating
terms but still keeps the simple mathematical structure of the
ordinary RWA. We have investigated the energy levels of ground and
lower-lying excited states, as well as the time-dependent quantum
dynamics. It is obvious that the approach is quite simple and can be
easily extended to more complicated situation. Besides, the results
are compared with the numerically exact ones to show that for weak
and intermediate coupling and moderate detuning our analytic
calculations are quantitatively in good agreement with the exact
ones.

Generally speaking, the generator (6) is for a displacement
transformation and the parameter $\xi$ may be used to indicate the
amount of the displacement of the oscillator. If $\xi=0$, that is,
without the transformation, we have the ordinary RWA (Eq.(5))
which is good for weak coupling and near resonance. If $\xi=1$,
that is, the oscillator can follow adiabatically the tunnelling of
the two-level system, we can have the generalized RWA of
Ref.\cite{iri3} which is good for higher frequency $\omega\gg
\Omega$ and strong coupling $g/\omega\gg 1$. Our $\xi$ is
determined in Eq.(12), $0<\xi<1$, which is in between the ordinary
RWA and the generalized RWA. Physically, our $\xi$ is to take into
account the nonadiabatic effect when the retardation of the
interaction between the two-level system and the quantum
oscillator with moderate detuning and intermediate coupling is
important. In addition, the bare coupling $g$ in original
Hamiltonian $H$ is replaced by the renormalized coupling
$g'=2g\eta\Omega/(\omega+\eta\Omega)$ in $H_{TRWA}$ because the
effect of the counter-rotating terms has been included.

\vskip 0.5cm

{\noindent {\large {\bf Acknowledgement}}}

This work was supported by the China National Natural Science
Foundation (Grants No.90503007 and No.10734020).

\newpage

\begin{center}
{\Large \bf Figure Captions }
\end{center}

\vskip 0.5cm

\baselineskip 20pt

{\bf Fig.1}~~~The energy levels of the ground state ($E_g$) and the
lower-lying excited states ($E_1$, $E_2$ and $E_3$, from bottom to
top) as functions of the ratio $g/\omega$. $\Omega=1$ and
$\omega=0.5$. Our results (dashed lines) are compared with the
numerically exact ones (solid lines) and those of the GRWA
Ref.\cite{iri2} (dashed-dotted lines).

\vskip 0.5cm

{\bf Fig.2}~~~The energy levels of the ground state ($E_g$) and the
lower-lying excited states ($E_1$, $E_2$ and $E_3$, from bottom to
top) as functions of the ratio $g/\omega$. $\Omega=0.5$ and
$\omega=1$. Our results (dashed lines) are compared with the
numerically exact ones (solid lines) and those of the GRWA
Ref.\cite{iri2} (dashed-dotted lines).

\vskip 0.5cm

{\bf Fig.3}~~~Time evolution $P(t)$ for $\Omega=1$, $\omega=0.5$ and
$g=0.4$. The dashed line is our transformed RWA calculation. The
solid line and the dashed-dotted line are the numerically exact
result and the ordinary RWA one, respectively.

\vskip 0.5cm

{\bf Fig.4}~~~Time evolution $P(t)$ for $\Omega=0.5$, $\omega=1$ and
$g=0.5$. The dashed line is our transformed RWA calculation. The
solid line and the dashed-dotted line are the numerically exact
result and the ordinary RWA one, respectively.

\vskip 0.5cm

{\bf Fig.5}~~~Time evolution $P(t)$ for $\Omega=0.25$, $\omega=1$
and $g=1$. The dashed line is our transformed RWA calculation. The
solid line and the dashed-dotted line are the numerically exact
result and the ordinary RWA one, respectively.

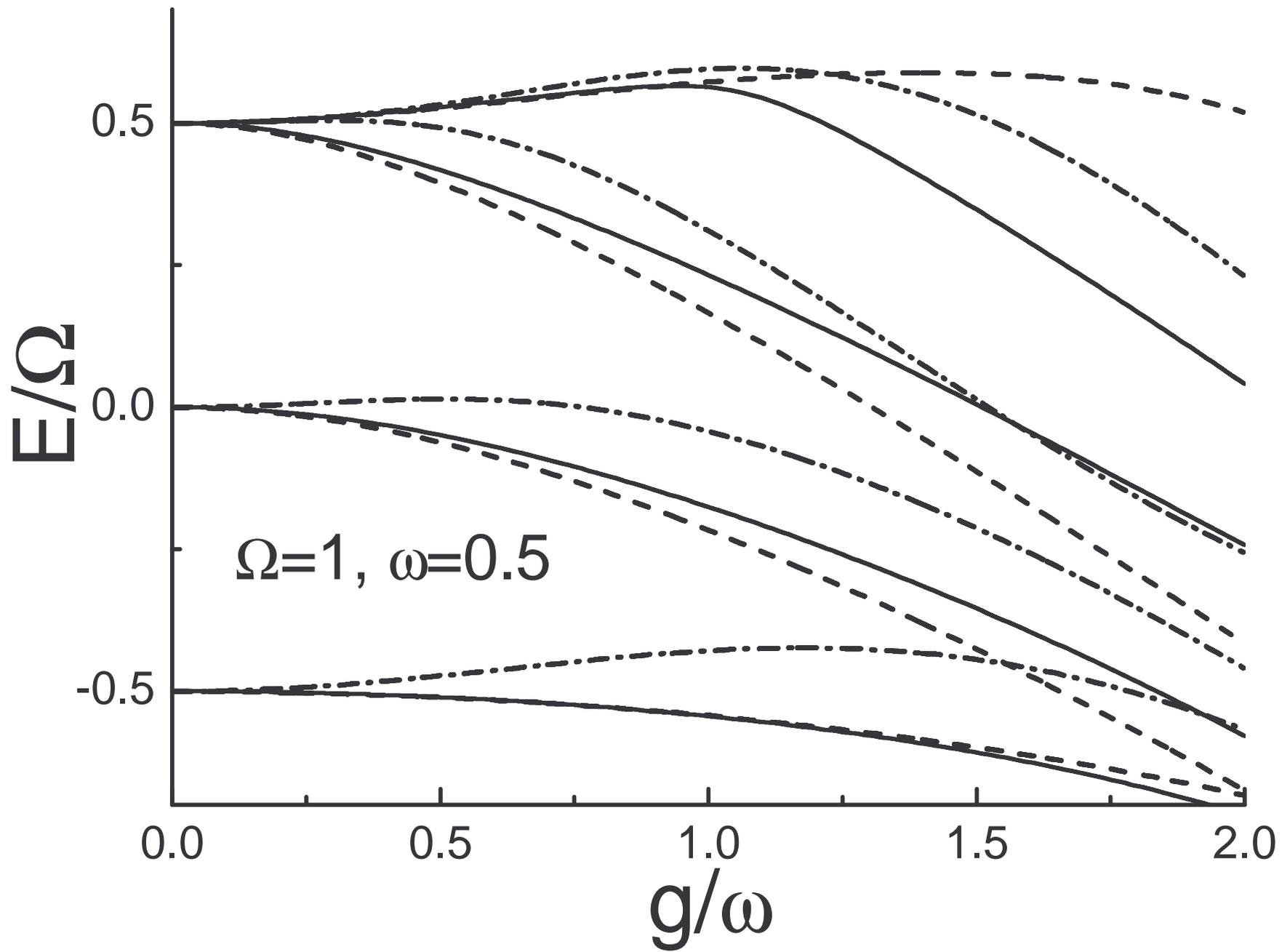
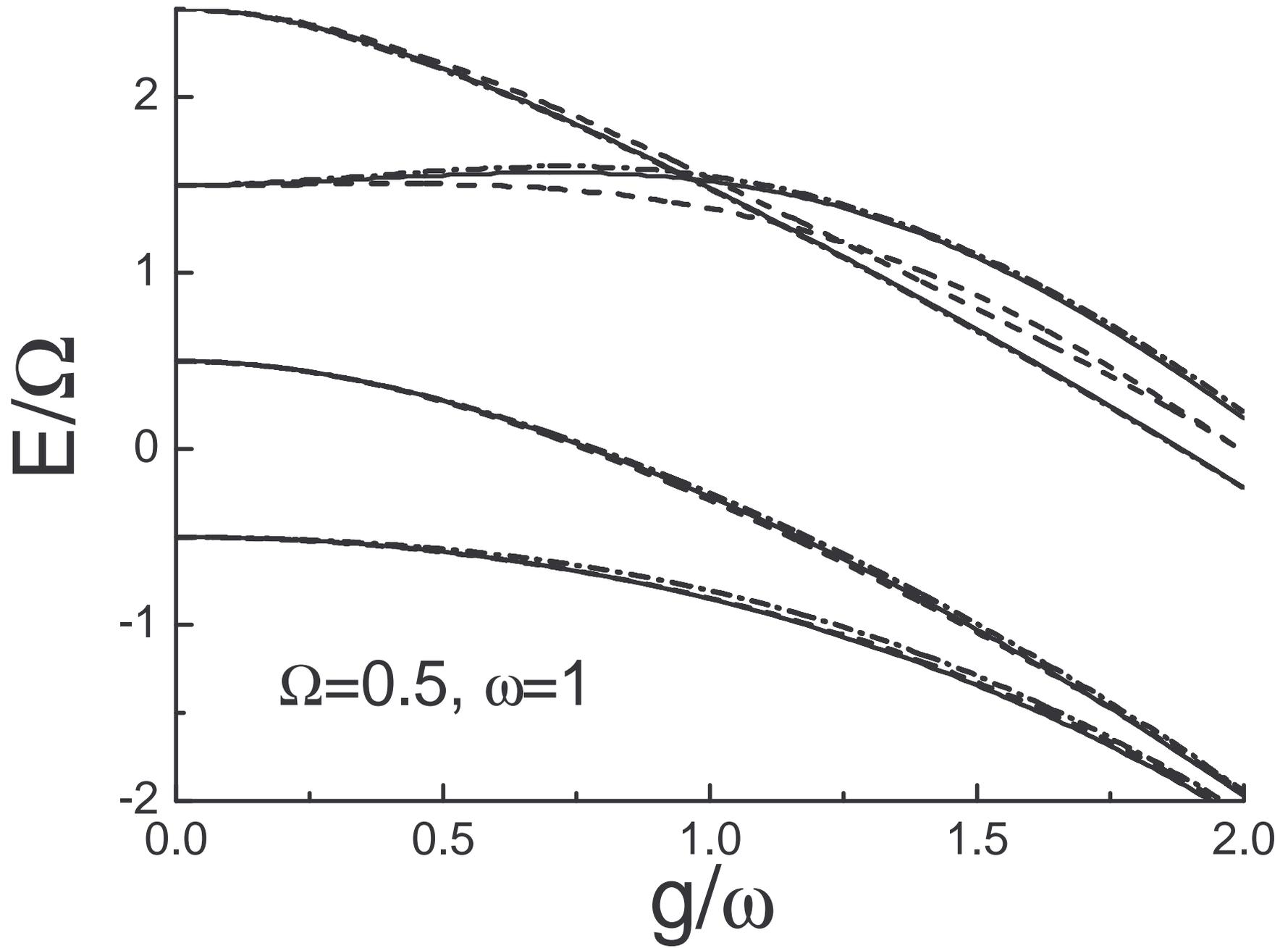
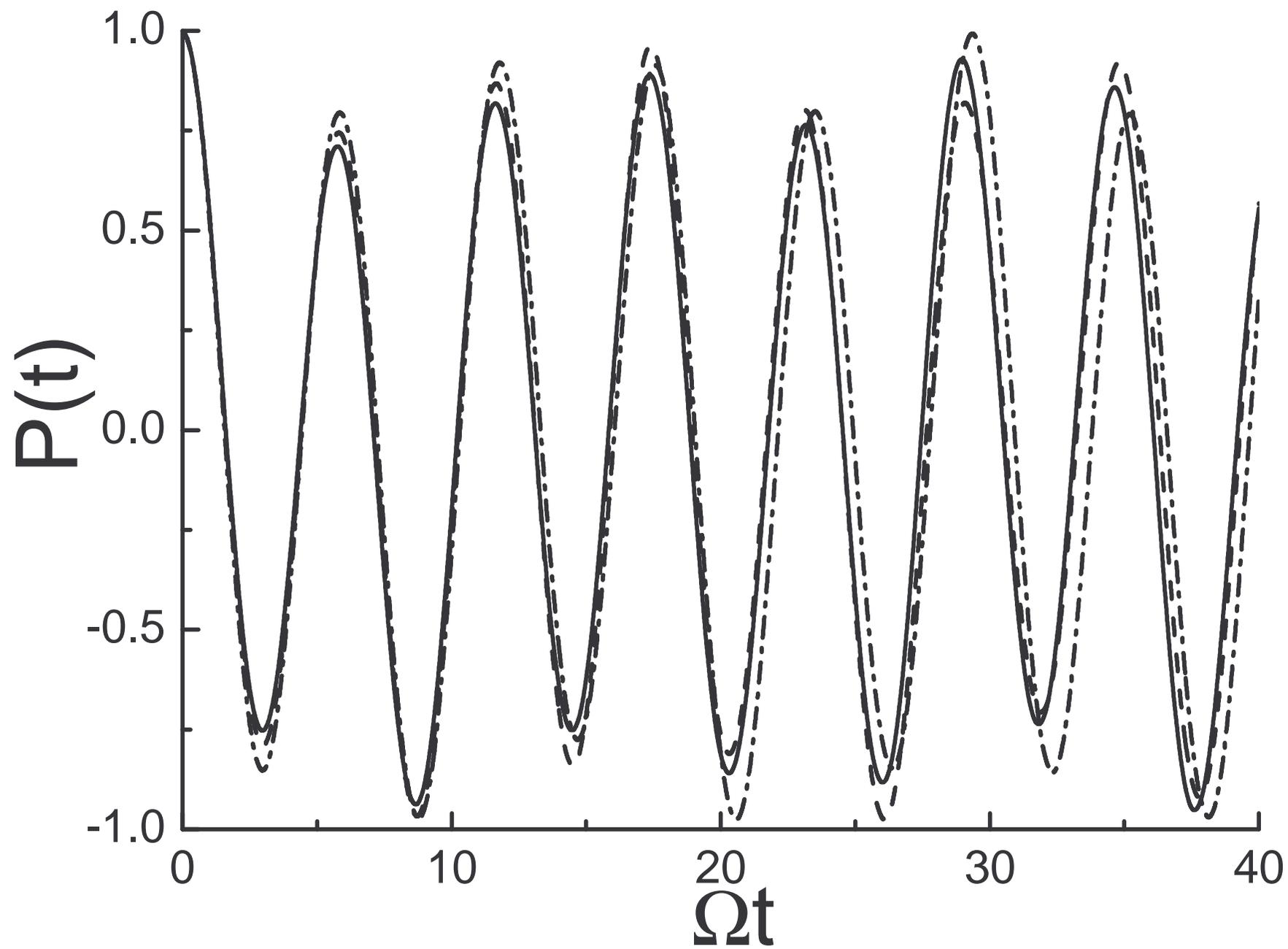
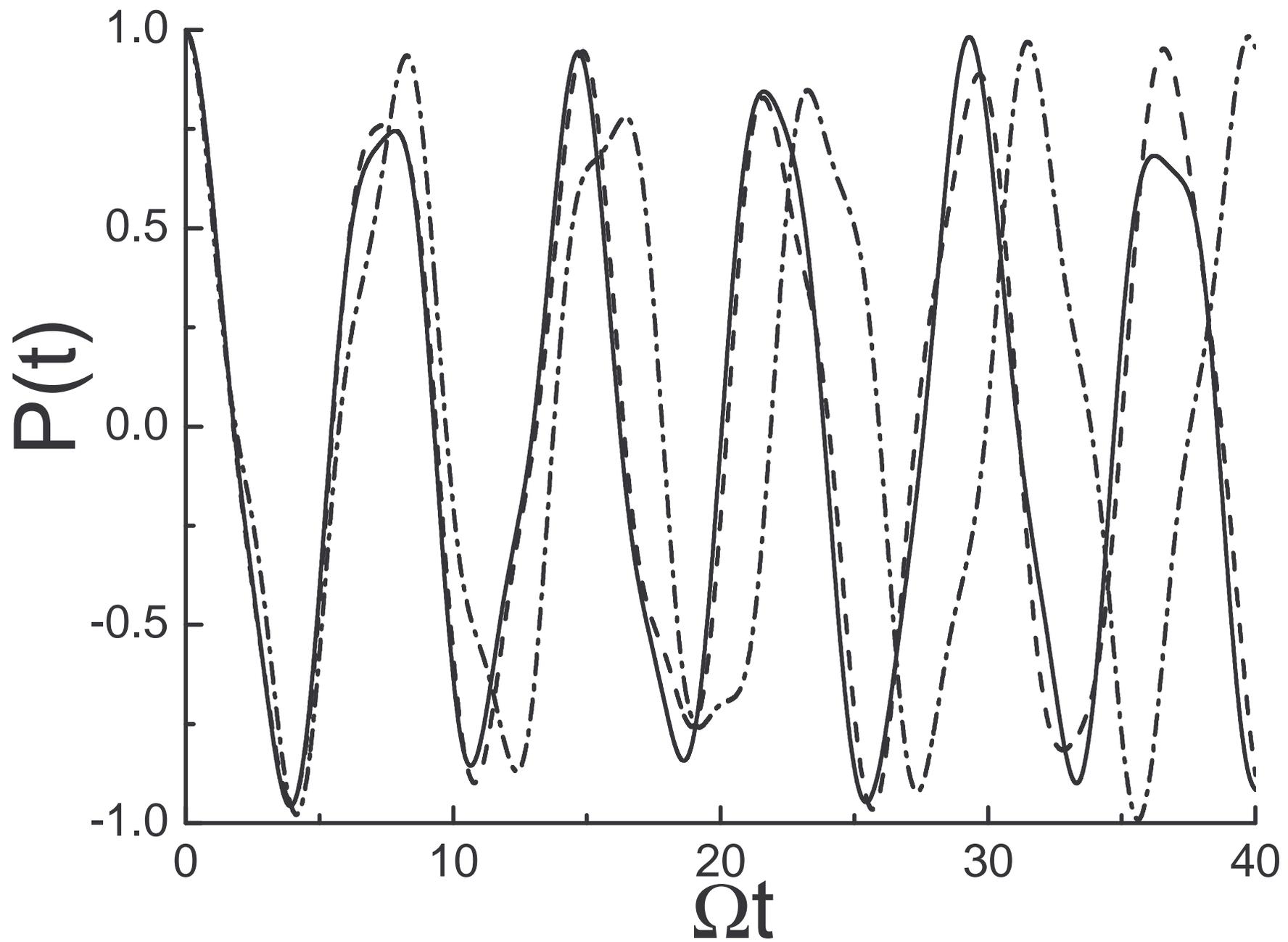
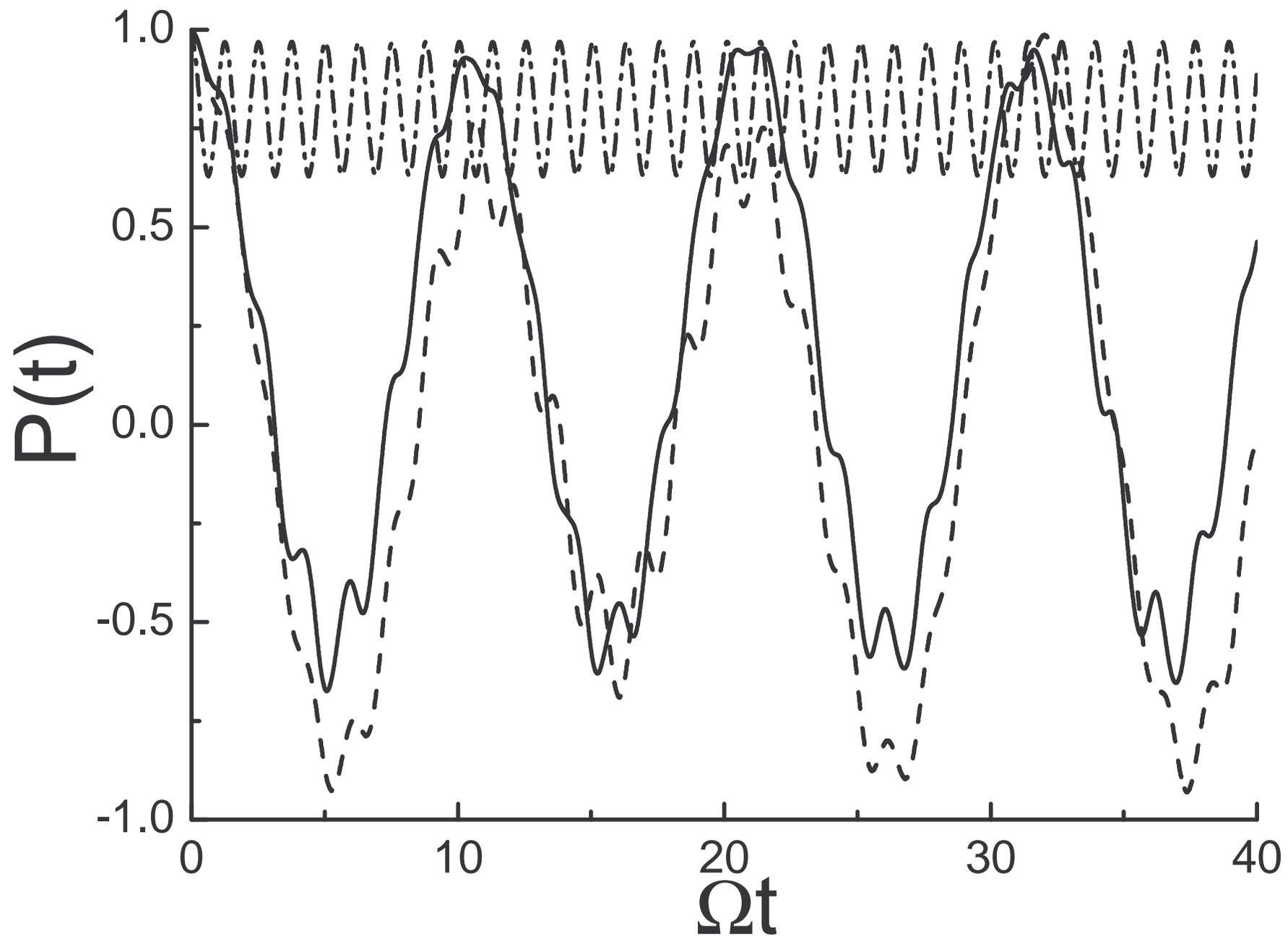
\end{document}